\documentstyle[11pt]{article}
\input{psfig.sty}
\begin{document}	
\pagestyle{plain}
\hsize = 6.0 in 				
\vsize = 8.0 in		   
\hoffset = -0.55 in
\voffset = -0.85 in
\baselineskip = 0.24 in	
\def\vecA{{\bf A}}
\def\vecB{{\bf B}}
\def\vecF{{\bf F}}
\def\vecJ{{\bf J}}
\def\vecX{{\bf X}}
\def\vecf{{\bf f}_{irr}}
\def\vecxi{\hbox{\boldmath $\xi$}}
\def\wt{\widetilde}

\vskip 0.4cm \noindent
\centerline{\large{\bf Equations for Stochastic Macromolecular Mechanics of}}
\vskip 0.25cm \noindent
\centerline{\large{\bf Single Proteins: Equilibrium Fluctuations, Transient}}
\vskip 0.25cm \noindent
\centerline{\large{\bf Kinetics and Nonequilibrium Steady-State}}
\vskip 0.75cm \noindent
\centerline{Hong Qian}
\vskip 0.3cm \noindent
\centerline{Department of Applied Mathematics}
\centerline{ University of Washington, Seattle, WA 98195, U.S.A.}
\centerline{(206)-543-2584 (Phone), (206)-685-1440 (Fax)}
\centerline{qian@amath.washington.edu}
\vskip 1.0cm  \noindent
\centerline{\today}
\vskip 1.0cm  \noindent

\centerline{\bf ABSTRACT}
\vskip 0.3cm

\begin{quote}
	A modeling framework for the internal conformational 
dynamics and external mechanical movement of single biological 
macromolecules in aqueous solution at constant temperature is 
developed.  Both the internal dynamics and external movement are 
stochastic; the former is represented by a master equation 
for a set of discrete states, and the latter is described by 
a continuous Smoluchowski equation.  Combining these two 
equations into one, a comprehensive theory for the Brownian
dynamics and statistical thermodynamics of single 
macromolecules arises.  This approach is shown to have 
wide applications.  It is applied to protein-ligand 
dissociation under external force, unfolding of 
polyglobular proteins under extension, movement along linear 
tracks of motor proteins against load, and enzyme catalysis by 
single fluctuating proteins.  As a generalization of the
classic polymer theory, the dynamic equation is capable of 
characterizing a single macromolecule in aqueous solution, in 
probabilistic terms, (1) its thermodynamic equilibrium with 
fluctuations, (2) transient relaxation kinetics, and most 
importantly and novel (3) nonequilibrium steady-state with heat 
dissipation.  A reversibility condition which guarantees an 
equilibrium solution and its thermodynamic stability is established, 
an {\em H}-theorem like inequality for irreversibility is obtained, 
and a rule for thermodynamic consistency in chemically pumped
nonequilibrium steady-state is given. 
\end{quote}
\vskip -0.5cm

\line(1,0){400}

\vskip 0.5cm

\vskip 1.0cm \noindent
\centerline{\bf I. Introduction}
\vskip 0.5cm \noindent

      Progress in optics, electronics, and computer science has
now made it possible to study biological macromolecules
in aqueous solution at constant temperature by observing 
experimentally and measuring quantitatively the behavior 
of single biological macromolecules. These studies have 
been providing and will continue to yield important information on the
behavior and properties of individual biomolecules and to reveal molecular
interactions in and the mechanisms of biological processes. The impact
which single-molecule studies will have on molecular biology may be gauged
by comparison with the pioneering studies on single channel proteins in
membranes, which have revolutionized physiology.$^1$
The highly quantitative data obtained in these novel measurements,
with piconewton and nanometer precision on the forces and movements of 
single macromolecules, complement those from traditional kinetic 
studies and structural determinations at the atomic level.  

	The novel experimental approach requires a consistent theoretical
framework for quantitatively understanding, interpreting, and integrating 
laboratory data.$^{2-6}$  The objective is to develop a unifying molecular 
theory, with thermodynamic consistency, which is capable of integrating 
the three classes of quantitative measurements on macromolecules: 
macroscopic (spectroscopic kinetics), mesoscopic (single molecules), 
and microscopic (atomic structures).  In this paper, we show how 
the spectroscopically defined kinetics, 
expressed in terms of discrete conformational states, is integrated with 
the mechanics of a macromolecule.  The philosophy behind this approach,
{\it Stochastic Macromolecular Mechanics}, is that we realize
the impracticality of representing the entire conformational space 
of a macromolecule with a high-dimensional energy function. 
Hence we still rely on a discrete-state Markov model with 
experimentally defined ``states'' and kinetic parameters.  However, 
we introduce a continuous energy landscape when there are
relevant, mechanical data on the single molecule.  Therefore the 
stochastic macromolecular mechanics is a mix of the discrete-state 
Markov kinetics with Brownian dynamics based on available and relevant 
experimental measurements.  It is a mesoscopic theory with a single set of 
equations. The theoretical approach helps researchers to identify the 
relevant (random) variables and key parameters in a macromolecular 
system or process, and provides them with the necessary equations. 

	The discrete-state master equation approach has been 
long accepted as the natural mathematical description for 
biochemical kinetics of individual molecules.$^{7,8}$
With more detailed information on molecular structures 
and energetics, the Smoluchowski's continuous description of 
overdamped Brownian dynamics has found numerous applications in 
condensed matter physics, polymer chemistry, and biochemistry of 
macromolecules$^{9,10}$.  H\"{a}nggi et al.$^{11}$ have
reviewed the related work with Kramers' approach to chemical rate
theory in which the assumption on overdamping is not warranted. 
However, for biological macromolecules in aqueous solution, and with
the time scale of biological interests, this assumption is generally
acceptable.  In dealing with a single protein molecule, the discrete 
approach is appropriate for spectroscopic studies$^3$ while the 
continuous approach is necessary for mechanical measurements.  
By combining these two descriptions, the stochastic macromolecular 
mechanics treats the internal conformational dynamics of proteins 
as well as its external mechanics.  In particular, both internal 
and external forces are explicitly considered.  On the 
mathematical side, such a combination leads to coupled stochastic 
processes,$^{12}$ giving rise to three different classes
of problems: reversible stationary processes (in a physicist's term,
thermal equilibrium with fluctuations), nonstationary processes 
(kinetic transient), and irreversible stationary processes 
(nonequilibrium steady-state with dissipation).  The last class of 
processes is novel$^{13-15}$ and necessary for modeling 
motor protein (e.g. kinesin and myosin) movement and 
energetics,$^{16-26}$ as well as other ``active macromolecules''.

	The differential equations in stochastic macromolecular
mechanics are Fokker-Planck-master type (linear diffusion-convection
equations with variable coefficients) based on conservation of 
probability.$^8$  This type of equations is different from 
the well-studied nonlinear reaction-diffusion equations 
$^{27,28}$ with distinctly different 
mathematical properties.$^{15,21}$

	We would like to point out that there is already a large 
literature on both Smoluchowski equation and master equations.  The 
novelty of our formalism is that (i) in order to combine the two 
types of equation into one, a condition analogous to the 
``thermodynamic box'' in the elementary chemical kinetics needs to
be introduced.  This relation, called potential condition,$^{29}$
yields a constraint between the energy functions in the Smoluchowski 
equation and the transition rates in the master equation.  This 
constraint guarantees the time-reversibility of the stationary 
solution to the stochastic macromolecular mechanics (SM$_3$), similar 
to that of fluctuation-dissipation relation required in modeling
equilibrium Brownian motion.  More importantly, however, is that
(ii) for single macromolecules like motor proteins, this condition
is violated due to the presence of chemical pumping (i.e.,
ATP hydrolysis).$^{21,22}$ This latter class of models based on 
Smoluchowski-master equation is relatively new. Its relation to
the phenomena of stochastic resonance has been revealed only
recently.$^{30,31}$  And finally, (iii) the stationary solution
to the SM$_3$ with chemical pumping defines a nonequilibrium 
steady-state the thermodynamics of which can be rigorously 
investigated.  By thermodynamics, we mean the entropy production,
heat dissipation, nonlinear irreversible force-flux relationship, 
and the law of thermodynamics.  It is this third aspect of the 
SM$_3$, we suggest, makes the Smoluchowski approach more powerful
and fundamental than that is generally aware. SM$_3$ as a 
statistical thermodynamic theory for macromolecules in isothermal 
aqueous solution, both passive and active, will have wide
applications, and deserves further 
investigations in the light of single-molecule experiments. 

	A second objective of this paper is introducing the 
physical chemists who are interested in the Smoluchowski's
approach to the unique and exciting opportunity in 
the current biophysics of nonequilibrium macromolecules on 
the level of single molecules.
In that sense, SM$_3$ is a generalization of the classic
polymer theory$^9$ into the nonlinear (Sec. II) and nonequilibrium
(Sec. III) regime.

	The paper is organized with increasing complexity as follows.  
In Sec. II we show how external force is introduced into the kinetics 
of protein-ligand dissociation.   We point out that mechanical 
measurements on single molecular complex depend critically
on the experimental conditions - the stiffness of the force probe 
and the rate of its retraction.  In Sec. III, polymers consisting 
of nonlinear subunits are introduced and we show how the internal 
kinetics is coupled to the external mechanics and movement. A
Boltzmann's {\em H}-theorem like inequality is obtained; and the
importance of nonlinear spring in serial leading to complex 
mechanical behavior is discussed.  Sec. IV introduces 
the ATP hydrolysis into the model.  Detailed balance, nonequilibrium 
steady-state, and thermodynamic consistency are discussed.  Sec. V. 
shows how SM$_3$ can be applied to the well-studied problem of 
fluctuating enzyme and yields new insights.  In particular, we show
how the classic concepts such as thermodynamic linkage and induced 
fit are consequence of the detailed balance, and can be 
generalized and quantified.  A summary is given in Sec. VI.

\vskip 1.0cm \noindent
\centerline{\bf II. Macromolecular Mechanics of Protein-Ligand Dissociation}
\vskip 0.5cm \noindent

	In this section, we discuss the dissociation of a
single protein-ligand complex under an external force introduced by
an experimenter.$^{32-34}$ As in any mechanics measurement, one first
is interested in the position of the ligand with respect to 
the center of the mass of the protein.  The next mechnical 
quantity is what is the forces acting on the ligand.  This leads
to a Newtonian equation in which one neglects the acceleration term
\begin{equation}
       \beta \frac{dx}{dt} = F_{int}(x) + F_{ext}(x,t) + f(t). 
\label{sde}
\end{equation} 
The four terms are $(i)$ frictional force with frictional coefficient
$\beta$, $(ii)$ intermolecular force between the ligand and the 
protein, with potential energy function $U_{int}(x)$: 
$F_{int}(x)$=$-dU_{int}(x)/dx$, $(iii)$ the external force, and 
$(iv)$ the stationary, pure random force due to collisions between 
the ligand and the solvent molecules: $\langle f(t)\rangle=0$.  
Because of the presence of the random force $f(t)$, the movement 
$x(t)$ is stochastic, e.g., it is a Brownian motion.  Mathematically 
equivalent, the Smoluchowski's description of overdamped Brownian 
dynamics is based on a partial differential equation of 
parabolic type:$^{29,35}$  
\begin{equation}
  \beta\frac{\partial P(x,t)}{\partial t} = k_BT
   \frac{\partial^2 P(x,t)}{\partial x^2}-
   \frac{\partial}{\partial x}\left[(F_{int}(x)+F_{ext}(x,t)) P(x,t) \right]
\label{Smo}
\end{equation}  
where $P(x,t)$ is now the {\bf probability density} of the ligand 
being at $x$ at time $t$.  $k_B$ is the Boltzmann constant, and
$T$ is temperature which characterizes the magnitude of the 
random force $f(t)$: $\langle f(t)f(t')\rangle=2k_BT\delta(t-t')$.   

	The above Eq. (\ref{sde}) and (\ref{Smo}) lay the 
mathematical basis for all models, but the choices for 
$U_{int}(x)$ and $F_{ext}$ set the difference between 
different models.  In the work of Shapiro and Qian,$^{36-38}$
$U_{int}=V_0[(x_0/x)^{12}-2(x_0/x)^6]$ with a smooth repulsive force,
and $F_{ext}=k(x-d(t))$ where $k$ is the stiffness of the force 
probe exerting the external force, and $d(t)=vt$ is the 
position of the piezoelectric motor which drives the force probe,
$v$ is the retracting velocity.  In the work of Evans and 
Ritchie,$^{39,40}$ $U_{int}=-V_0(x_0/x)^n$ $(x>x_0)$ with an 
abrupt repulsion at $x_0$, and $F_{ext}=-F(t)$ is independent 
of $x$.  These differences give qualitatively similar but 
qualitative different results.  Hence they can be quantitatively
tested against experimental date. 

	Fig. 1 shows the results of simulations on the 
force-displacement curve for a protein-ligand complex
with a simple 6-12 Lennard-Jones potential, measured using an
elastic force probe.  It is important to note that all the
differences between the curves are due to the difference in the 
stiffness of the force probe ($k$), the rate of retraction ($v$), 
and the temperature ($k_BT$) at which the measurements are 
carried out. Therefore, this calculation demonstrates that the ``raw'' 
experimental data can only be understood, in general, in terms 
of a molecular model.  It is important to realize the 
significance of the measurement aparatus on the experimental 
data.

\vskip 1.0cm \noindent
\centerline{\bf III. Macromolecular Mechanics of Polyglobular Protein
Unfolding}
\vskip 0.5cm \noindent

	In the previous section on protein-ligand dissociation, we
have completely neglected the conformational change within the protein
itself.  The protein was treated as a rigid body exerting a force on
the ligand.  A more realistic model must consider the possibility
of the protein's internal conformational change due to the external force, 
acting via the ligand.  In this section, we study the unfolding of a
polyglobular protein under extensional force.  This problem naturally
involves the internal dynamics of the macromolecules. 

	To be concrete, let's consider the recent experimental
work on giant muscle protein titin.$^{41-43}$  Titin is a protein
with many globular domains (subunits) in serial.  The subunits
unfold under an external force pulling the entire molecule. 
The folded state of each subunit is rigid, and the unfolded state
of each subunit can be regarded as a coiled polymer spring.  Hence
the conformational state of the entire protein, to a first order
approximation, can be characterized by $n$: the number of unfolded
subunits within the molecule.  Let's assume the total number of 
subunits are $N$, and let $x$ be the total extension of the 
titin molecule (along the axis of external force), then a realistic
characterization of a titin molecule is by two dynamic variables 
$(x,n)$, $(x\ge 0,\ 0\le n\le N)$. 

	The equation of motion for $x$ is again
\begin{equation}
   \beta\frac{dx}{dt} = -\frac{dU_{int}(x,n)}{dx} + F_{ext} + f(t)
\label{titin1}
\end{equation}   
in which $n$ is itself a random (discrete-state Markov) process.  
Hence the above equation is coupled to a master equation
\[ \frac{\partial P(n,t)}{\partial t} = 
     (N-n+1)\lambda_u(x,n-1)P(n-1,t) +(n+1)\lambda_f(x,n+1)P(x,n+1,t) \] 
\begin{equation}
    -[n\lambda_f(x,n)+(N-n)\lambda_u(x,n)]P(n,t)  
\label{titin2} 
\end{equation} 
where $P(n,t)$ is the probability for $n$ at time $t$.
$\lambda_f$ and $\lambda_u$ are folding and unfolding rate constants
of individual subunits.  They are functions of the force acting on the 
subunit, which in turn is determined by the total extension of the 
molecule ($x$) and the number of unfolded domains in the chain ($n$). 

	A comprehensive description of both the internal dynamics
and external movement can be obtained by combining Eq. (\ref{titin1}) 
and (\ref{titin2}).  We therefore have 
\begin{eqnarray}
  \frac{\partial P(x,n,t)}{\partial t} &=& \left(\frac{k_BT}{\beta}\right)
   \frac{\partial^2 P(x,n,t)}{\partial x^2} + \frac{1}{\beta}
   \frac{\partial}{\partial x} \left[\left(\frac{dU_{int}(x,n)}{dx}
		+F_{ext}\right)
                            P(x,n,t) \right]   \nonumber\\
   &+& (N-n+1)\lambda_u(x,n-1) P(x,n-1,t)    \nonumber\\
   &+& (n+1)\lambda_f(x,n+1) P(x,n+1,t)    \nonumber\\
   &-& [n\lambda_f(x,n)+(N-n)\lambda_u(x,n)]P(x,n,t) 
\label{titin3}  
\end{eqnarray}
where $P(x,n,t)$ is the {\bf joint probability distribution} 
of the titin molecule having internally $n$ unfolded domains
and external extension $x$. 

	As in the previous section, particular models will 
provide specific $\lambda_f(x,n)$, $\lambda_u(x,n)$, and 
$U_{int}(x,n)$.  These functions are not totally independent,
however.  Microscopic reversibility dictates that
\begin{equation}
  \frac{\lambda_u(x,n)}{\lambda_f(x,n+1)}
        = \frac{n+1}{N-n} 
        \exp\left[-\frac{U_{int}(x,n+1)-U_{int}(x,n)}{k_BT}\right].
\label{pc}
\end{equation} 
This condition guarantees that the stationary solution to 
Eq. (\ref{titin3}) is a thermodynamic equilibrium with time 
reversibility.  Furthermore, this {\bf reversibility condition} 
(also known as potential condition and detailed balance) also 
guarantees the thermodynamic stability of the equilibrium state
in terms of a generalized free energy function.  
As we shall see below, without (\ref{pc}), 
the stationary solution in general represents a nonequilibrium 
steady-state with dissipation. 

	With the reversibility condition and in the absence of
external force $F_{ext}$, it is easy to verify that the stationary 
solution to Eq. (\ref{titin3}) is
\begin{equation}
	P^*(x,n) = Z^{-1}e^{-U(x,n)/k_BT}    
\end{equation}
where 
\[   Z  = \sum_{n=0}^N \int_0^{\infty}e^{-U(x,n)/k_BT}.    \]
The time-dependent solutions to (\ref{titin3}) are
dynamic models for nonstationary transient kinetics of 
macromolecules.  

	We now show that the equilibrium solution $P^*(x,n)$ is
asymptotically stable.  By stability, we mean a molecule approaches to 
its equilibrium state irrespective of its initial state.  
We introduce a free energy functional:
\begin{eqnarray*}
    \Psi[P(x,n,t)] &=& \sum_{n=0}^N\int_0^{\infty} 
    \left(U(x,n)P(x,n,t)
		+k_BTP(x,n,t)\ln P(x,n,t)\right)dx   \\
     &=& -k_BT\ln Z + k_BT\sum_{n=0}^N\int_0^{\infty} P(x,n,t)\ln
	\left(\frac{P(x,n,t)}{P^*(x,n)} \right)dx 
\end{eqnarray*} 
in which the second term (known as relative entropy$^{44,45}$)
is always nonnegative and equal to zero 
if and only if $P(x,n,t)=P^*(x,n)$.$^{29}$  Based on Eq. (\ref{titin3})
the time derivative of $\Psi$ is 
\begin{equation}
   \dot{\Psi}[P(x,n,t)] = -\sum_{n=0}^N \int_0^{\infty} 
        \left[\beta^{-1}J^2(x,t)P^{-1} 
	+ k_BT\left(J_n^+-J_n^-\right)
	\ln\left(\frac{J_n^+}{J_n^-}\right)\right] dx 
       \le 0
\label{frate} 
\end{equation} 
where
\[    J(x,t) = -k_BT\frac{\partial P(x,n,t)}{\partial x}
               - \frac{dU(x,n)}{dx} P(x,n,t)    \]
and
\[   J_n^+(t) = (n+1)\lambda_f(x,n+1)P(x,n+1,t), \hspace{0.3cm}
     J_n^-(t) =	(N-n)\lambda_u(x,n)P(x,n,t).             \]
The integrand in Eq. (\ref{frate}) is always positive.  Hence
$\Psi$ is a Lyapunov functional for the time-dependent solution 
of Eq. (\ref{titin3}), which guarantees $P^*$ to be
asymptotically stable.  Furthermore, because of the reversibility
condition, the differential equation in (\ref{titin3}) is
symmetric and all its eigenvalues are real, indicating 
relaxations in such systems can not oscillate. 

	The physical interpretation of the above result is 
important: It relates the equation of SM$_3$ to the second
law of thermodynamics. The $\Psi$-function is the generalization
of the equilibrium free energy of a closed, isothermal molecular 
system.  $\Psi$ decreases monotonically to its minimum $-k_BT\ln Z$, 
the Gibbs free energy, when the system reaches its equilibrium.  
It is intriguing to note that the dynamics in Eq. (\ref{titin3}) is 
not governed by the gradient of the free energy.  Nevertheless, one 
should see the analogy between Eq. (\ref{frate}) and the {\em H}-theorem of 
Boltzmann in his approach to irreversibility in an isolated
system (microcanonical ensemble).

	To analyze the stochastic dynamics of a complex macromolecule
under extensional force, it is important first to have an essential 
understanding of its nonlinear mechanical property.  A polyglobular 
protein model is a generalization of the classic ``bead-and-spring'' 
to nonlinear spring.$^{9}$  The protein subunits all have {\it two}
energy minima while a simple Hookean spring has only one.  This
leads to fundamentally different behavior of the macromolecule. 
To illustrate this, let's apply the elementary Ohm's law for 
nonlinear springs in serial: {\it the force on the springs are 
the same while the displacement is additive.}
For simplicity, assume each subunit has a potential function 
(and force) given in Fig. 2.  This energy function has been used 
in recent work on globular protein folding kinetics.
Then Fig. 3 gives a quantitative force-extension curve expected 
from a polyglobular protein with three subunits.  As one can
see, the most striking feature is the possibility of multiple
branches of the curve with a given force.  This is due to the
combinatorics shown in Table I.

\vskip 1.0cm \noindent
\centerline{\bf IV. Macromolecular Mechanics of Motor Protein Movement}
\vskip 0.5cm \noindent

	With the presence of the reversibility (potential) condition, 
the previous model represents a ``passive'' complex molecule.  Without 
the external force, such molecules relax, multi-exponentially, to a 
thermodynamic equilibrium.  They are biochemically interesting, 
but they are not ``alive''.  One
could argue that one of the fundamental properties of a living
organism is the ability to convert energy among different 
forms (solar to electrical, electrical to chemical, chemical 
to mechanical, etc.). We now show how stochastic macromolecular
mechanics can be used to develop models for chemomechanical
energy transduction$^{16,17}$ in single motor proteins.$^{46}$
In the absence of an external force, a motor protein is
undergoing nonequilibrium steady-state with ATP hydrolysis and
generating heat -- representing a rudimentary form of energy 
metabolism.   

       The key for developing a theory for motor protein is to consider 
that while biochemical studies of a protein in test tubes probe a
set of discrete conformational states of the molecule, the
mechanical studies of a protein measure positions and forces.
Internally, a motor protein has many different conformational states 
within a hydrolysis cycle, and a reaction scheme usually can be 
deduced from various kinetic studies. While the protein is 
going through its conformational cycles, 
its center of mass moves along its designated linear 
track (e.g., kinesin on a microtubule, myosin on an actin filament, 
and polymerase on DNA) which usually has a periodic structure.
The movement is stochastic; the interaction 
between the motor and the track (the force field) are usually 
different for different internal states of the molecule.

	These basic facts lead to the following equation
\begin{eqnarray}
 \frac{\partial P(x,n,t)}{\partial t} &=& \left(\frac{k_BT}{\beta}\right)
   \frac{\partial^2 P(x,n,t)}{\partial x^2}+\frac{1}{\beta}
   \frac{\partial}{\partial x}
        \left[\left(\frac{dU_{int}(x,n)}{dx}+F_{ext}\right) 
		P(x,n,t) \right]   \nonumber\\
    &+& \sum_{k=1}^N \left[\lambda_{kn}(x)P(x,k,t)-
 		\lambda_{nk}(x)P(x,n,t)\right],
                 \label{motor1}
\end{eqnarray}
where $P(x,n,t)$ denote the joint probability of a motor
protein with internal state $n$ and external position $x$.
$U_{int}(x,n)$ is the interaction energy between the protein
in state $n$ and the track.  $\lambda_{\ell m}(x)$ is the 
transition rate constant from internal state $\ell$ to 
state $m$ when the protein is located at $x$. 
Some of the $\lambda$'s are pseudo-first order rate constants which
contain the concentrations [ATP], [ADP], and [Pi]. 

	When the ratio [ADP][Pi]/[ATP] = $K_{eq}$, the equilibrium 
constant for the hydrolysis reaction, the $U(x,n)$ and 
$\lambda_{\ell m}^*(x)$ are again constrained by the reversibility 
(the superscript $^*$ is to indicate that the pseudo-first order 
rate constants are calculated in terms of the equilibrium 
concentrations):
\begin{equation}
    \frac{\lambda_{\ell m}^*(x)}{\lambda_{m\ell}^*(x)} =
   \exp\left(-\frac{U_{int}(x,m)-U_{int}(x,\ell)}{k_BT} \right).
\label{motor2}
\end{equation} 
This relation was called ``thermodynamic-consistency'' by 
T. L. Hill$^{47}$ 
in his landmark contribution to the Huxley's theory of muscle 
contraction$^{48}$.  It has to be satisfied by every motor protein 
models.  It is clear, however, that the ATP, ADP, and Pi 
can be kept at arbitrary values. Hence in general the stationary 
solution of the Eq.  (\ref{motor1}) will be a 
nonequilibrium steady-state. Sustaining the 
concentrations is a form of ``pumping'' which keeps the system
at nonequilibrium steady-state$^{31}$ with
positive entropy production and heat generation.$^{14,15}$  
Such a molecular device is also known as isothermal 
ratchet.$^{18,19,22}$ See Ref. 21 and 15 for reviews of the vast
literature.  If the concentrations are not actively sustained, 
then they will change slowly (since there is only a single 
molecule at work hydrolyzing ATP) and eventually reach a thermal 
equilibrium state in which (\ref{motor2}) is satisfied. 

	As in the previous sections, a practical model 
requires specific choices for the parameters $U$ and 
$\lambda$.  In the past several years, a large amount of 
work have appeared on modeling translational motor proteins 
such as myosin and kinesin$^{15,18,19,21,25,21}$ and 
rotational motor proteins such as ATP synthase.$^{49,50}$
A thermodynamically valid model for a motor protein has to
satisfy Eq. (\ref{motor2}) when [ADP][Pi]/[ATP] = $K_{eq}$,
but in general without a potential function.  This rule has 
not been enforced in some of the models. 

	Simpler but phenomenological models based on 
discrete-state kinetics have also been developed for 
motor protein kinetics and energetics.$^{20,23-25}$  It is
important to point out that these models are completely
in accord with the present continuous theory.  However,
drastic simplifications are used in order to make the models
more accessible to experimental data.  Fig. 4 shows
the conceptual relationship between these two classes of models.
Therefore, the discrete model should not be viewed as 
an alternative to the SM$_3$, rather it is a simplification 
which can be further scrutinized in terms of the general theory 
of SM$_3$.  The stistical thermodynamics associated with the 
continuous approach, however, can be developed in parallel
for the discrete models.$^{25}$

\vskip 1.0cm \noindent
\centerline{\bf V. Macromolecular Mechanics of Fluctuating Enzyme}
\vskip 0.5cm \noindent

	Equilibrium conformational fluctuation of proteins play an important role in enzyme kinetics.  The theory of 
fluctuating enzyme$^{51}$ can be developed naturally in 
terms of the above equations for stochastic macromolecular 
mechanics.  Let's consider a single enzyme, with its internal
conformation characterized by $x$, and $N$ number of 
substrate molecules.  The enzyme catalyzes a reversible
isomerization reaction between two forms of the substrate
(reactant and product), with rate constants $\lambda_+(x)$ 
and $\lambda_-(x)$.

	The equation for the catalytic reaction coupled with the 
enzyme conformational fluctuation, according to stochastic macromolecular
mechanics, is
\[ \frac{\partial P(n,x,t)}{\partial t}
   = -[n\lambda_++(N-n)\lambda_-]P(n,x,t) 	
		+ (n+1)\lambda_+P(n+1,x,t)   \]
\begin{equation}
     + (N-n+1)\lambda_-P(n-1,x,t)   
     +D\frac{\partial^2 P(n,x,t)}{\partial x^2}
     + k\frac{\partial}{\partial x} 
         (xP(n,x,t)), \hspace{0.3cm}
\label{sm3fe}
\end{equation}
\[ 		(0\le n \le N)   		\]
where $P(n,x,t)$ is the probability of at time $t$ having $n$ 
number of reactant molecules and the enzyme internal
conformation being at $x$.  $D$ and $k$ characterize 
the protein conformational fluctuation. $x$ is {\it perpendicular}
to the isomerization reaction coordinate as first proposed by 
Agmon and Hopfield,$^{10}$ in contrast to the other models which 
address random energy landscape {\it along} the reaction 
coordinate.$^{52}$ Eq. (\ref{sm3fe}), which is essentially the same 
equation for the modeling of polyglobular protein unfolding 
(Eq. \ref{titin3}), unifies and generalizes most of the important 
works on fluctuating enzymes.

	Along this approach, most work in the past have addressed the 
non-stationary, time-dependent solution to (\ref{sm3fe}).  These
studies are motivated by macroscopic experiments which are
initiated ($t=0$) with all the substrate in only the reactant form.
If $\lambda_-(x)=0$ and $N=1$, Eq. (\ref{sm3fe})
is reduced to that of Agmon and Hopfield.$^{10}$
If $\lambda_-(x)=0$ but $N$ is large, then one can introduce a
continuous variable $\xi=n/N$, known as the {\it survival
probability}, and Eq. (\ref{sm3fe}) can be approximated 
as (see Appendix I for more discussion)
\begin{equation}
   \frac{\partial P(\xi,x,t)}{\partial t}
   = \lambda_+(x)\frac{\partial}{\partial \xi}(\xi P(\xi,x,t))
     +D\frac{\partial^2 P(\xi,x,t)}{\partial x^2}
     + k\frac{\partial}{\partial x} 
         (x P(\xi,x,t)).
\label{sm3fe2}
\end{equation}
At $t=0$, Prob$\{\xi=1\}=1$. 

	The moments of $\xi$,
$\langle\xi^m\rangle(x,t)=\int_0^1\xi^mP(\xi,x,t)d\xi$,
can be easily obtained from 
Eq. (\ref{sm3fe2}):
\begin{equation}
 \frac{\partial\langle\xi^m\rangle}{\partial t}
     = -m\lambda_+(x) \langle\xi^m\rangle
       + D\frac{\partial^2 \langle\xi^m\rangle}{\partial x^2}
       + k\frac{\partial}{\partial x} 
            \left(x\langle\xi^m\rangle\right). 
\label{sm3fe3}
\end{equation}
Note Eq. (\ref{sm3fe}) with $N=1$ and Eq. (\ref{sm3fe3}) for
$\langle\xi\rangle$ are idential.
For $\lambda_+(x)=\alpha x^2+\beta x+\gamma$, Eq. (\ref{sm3fe3}) 
can be exactly solved by various methods if one realizes that
its solution has a Gaussian form$^{53-55}$ (also see Appendix II).  
From Eq. (\ref{sm3fe3}) one immediately 
sees that high-order moments $\langle\xi^m\rangle$ is related to 
$\langle\xi\rangle$ by $\lambda_+(x) \rightarrow m\lambda_+(x)$.$^{56}$

	Different choices for $\lambda_+(x)$ lead to quantitatively
different models for fluctuating enzymes in the literature. 
$\lambda_+(x)\propto e^{-\alpha x}$ represents a fluctuating 
activation energy barrier;$^{10}$ 
$\lambda_+(x)=\alpha (x+\overline{x})$ $(>0)$ representing a 
fluctuating cofactor concentration;$^{57}$ $\lambda_+(x)$
= $\alpha x^2$ representing a fluctuating geometric bottleneck.$^{53}$ 

	We now consider the reversible reaction 
(with $\lambda_-(x)\neq 0$) which has not been discussed previously.
This class of models is more appropriate for recent measurements in 
single-molecule enzymology.$^{3}$  Again we assume 
$N$ being large.  Hence we have 

\[ \frac{\partial P(\xi,x,t)}{\partial t}
   = \frac{\partial}{\partial \xi} \left( D_{\xi}(\xi,x)
     \frac{\partial P(\xi,x,t)}{\partial \xi} \right) +
     \frac{\partial}{\partial \xi} 
		\left(V_{\xi}(\xi,x)P(\xi,x,t)\right)    \]
\begin{equation}
     +D\frac{\partial^2 P(\xi,x,t)}{\partial x^2}
     + \frac{\partial}{\partial x} 
         (V_x(x,\xi) P(\xi,x,t))
\label{sm3rev}
\end{equation}
where $D_{\xi}(\xi,x)=[\xi \lambda_++(1-\xi)\lambda_-]/2N$ and
$V_{\xi}=\xi \lambda_+-(1-\xi)\lambda_-$.  Eq. (\ref{sm3rev}) is a 2D 
diffusion-convection equation similar to a continuous model we
proposed for motor protein movement.$^{21}$  One 
important consequence of this formulation and the reversibility
condition is realizing that 
conformational fluctuations of the enzyme, $V_x$ can
not be independent of the substrate.  This constitutes
the essential idea of {\it induced fit}$^{58-60}$ and
thermodynamic linkage!$^{61,57}$
For equilibrium fluctuation, 
again reversibility (i.e., potential condition) dictates:$^{21}$
\begin{equation}
	\frac{\partial V_x(x,\xi)}{\partial \xi} =
  D \frac{\partial}{\partial x} 
	\left(\frac{V_{\xi}(\xi,x)}{D_{\xi}(\xi,x)}\right) =
    \frac{4ND(\lambda_-\lambda_+'-\lambda_+\lambda_-')
		\xi(1-\xi)}{[\xi \lambda_++(1-\xi)\lambda_-]^2}
\end{equation}
where $\lambda' = d\lambda(x)/dx$.  Therefore,
\[     V_x(x,\xi) = -\frac{\xi}{(\lambda_+-\lambda_-)^2}
       +\frac{\lambda_++\lambda_-}{(\lambda_+-\lambda_-)^3}
		 \ln[\lambda_+\xi+\lambda_-(1-\xi)]    \] 
\begin{equation}
       + \frac{\lambda_+\lambda_-}{(\lambda_+-\lambda_-)^3
		[\lambda_+\xi+\lambda_-(1-\xi)]}
       + V_0(x)
\end{equation}
where $V_0(x)$ is an arbitrary function of $x$ but it is independent of
$\xi$.  As can be seen, if $\lambda_-<<\lambda_+$, then there is no 
requirement for $\xi$-dependent $V_x$.  

\vskip 1.0cm \noindent
\centerline{\bf VI. Conclusions}
\vskip 0.5cm \noindent

	Biological macromolecules are the cornerstone of molecular 
biology.  Mathematical modeling of biomolecular processes requires 
a comprehensive and thermodynamically consistent theoretical basis 
upon which quantitative analyses can be 
carried out and rigorously compared with experiments. 
In this paper, a formal theory, we call {\bf stochastic macromolecular
mechanics}, is presented.  The theory offers a 
dynamic equation for describing the internal kinetics as well as
external motion of macromolecules in aqueous solution at
constant temperature. Systematically applying this theory to 
various biomolecular processes will bring molecular biophysics 
closer to the standard of theoretical chemistry and physics.
At present time, Smoluchowski is well-known for its importance
in calculating the microscopic fluctuations of an isothermal 
equilibrium system.$^{8,9,29,35}$  It is less known that it can 
also be a cogent model for a macromolecules under chemical 
pumping.$^{18,19,30,21}$ What has not been appreciated is that 
this mesoscopic model also yields equilibrium and nonequilibrium
thermodynamics for the macromolecule.   Therefore, it deserves
the same status as that of Newton's for mechanics, 
Navier-Stokes' for fluid dynamics, 
Maxwell's for electrodynamics, Schr\"{o}dinger's for quantum mechanics,
and Boltzmann's for statistical mechanics of isolated systems.

\vskip 1.0cm \noindent
\centerline{\bf Acknowledgement}
\vskip 0.5cm \noindent

I like to dedicate this paper to Professor Joel Keizer.  Officially 
I have not worked with him.  However, he had been a mentor to me 
in the past 10 years because our common interests in nonequilibrium 
statistical mechanics and biophysics, and because of Terrell Hill.  
One can easily find his influence on my work presented here.  
He will be sorely missed.

\vskip 1.0cm \noindent
\centerline{\bf Appendix I}
\vskip 0.5cm \noindent

	Let's use the well-known linear death process$^{62}$
as an example to illustrate the
continuous approximation for the discrete model: 
\begin{equation}
     \frac{dP_n(t)}{dt} = -n\lambda P_n(t) 
			+ (n+1)\lambda P_{n+1}(t)  
\label{lbp}
\end{equation}
where $P_n(t)$ is the probability of survival population being
$n$ at time $t$.  The solution to this equation is well known$^{62}$
\[   P_n(t) = \frac{N!}{n!(N-n)!} e^{-n\lambda t}
		\left(1-e^{-\lambda t}\right)^{N-n}        \]
where $N$ is the total population at time $t=0$.  It is 
easy to show that the moments
\begin{eqnarray}
     \langle n(t) \rangle &=& Ne^{-\lambda t},   \nonumber \\[0.2cm]
    \frac{\langle n^2(t)\rangle}{\langle n(t)\rangle^2}
      &=& 1+\frac{1}{N}\left(e^{\lambda t}-1\right),  \label{dmom} \\[0.2cm]
    \frac{\langle n^3(t)\rangle}{\langle n(t)\rangle^3} &=& 
          1+\frac{3(N-1)}{N^2}\left(e^{\lambda t}-1\right)
           +\frac{1}{N^2}\left(e^{2\lambda t}-1\right). 
               \nonumber
\end{eqnarray}

	We now consider the continuous counterpart of 
(\ref{lbp}) with $\xi=n/N$:
\[   \frac{\partial P(\xi,t)}{\partial t} = 
     \frac{\partial}{\partial\xi} \left(\xi P(\xi,t)\right)   \]
which has solution
\[    P(\xi,t) = e^{\lambda t}
		\delta\left(\xi e^{\lambda t} -1\right)    \]
for initial condition $P(\xi,0)=\delta(\xi-1)$.
The moments for $\xi$ are
\begin{equation}
      \langle \xi^k(t) \rangle = e^{-k\lambda t}.  
\label{cmom}
\end{equation}
Comparing (\ref{dmom}) and (\ref{cmom}), we note that
the continuous approximation is valid when the $N$ 
is large and $t$ is small.  More precisely,
$\ln N\gg \lambda t$.

\vskip 1.0cm \noindent
\centerline{\bf Appendix II}
\vskip 0.5cm \noindent

	Let's consider the following equation
\[  \frac{\partial u}{\partial t} 
      = D\frac{\partial^2 u}{\partial x^2} 
      + k\frac{\partial}{\partial x} (xu)
      -(\alpha x^2 + \beta x + \gamma)u    \]
In Gaussian form$^{53}$ which is equivalent to 
path integral calculation$^{54}$
$u(x,t)$ = $[2\pi\sigma(t)]^{-1/2}$
$\exp\left\{\nu(t)-[x-\mu(t)]^2/2\sigma^2(t)\right\}$,
and equate coefficients of like order terms in $x$ we
have
\begin{eqnarray*}
   d\nu/dt &=& -\alpha[\mu^2(t)+\sigma^2(t)]-\beta\mu(t)-\gamma \\
   d\mu/dt &=& -[k +2\alpha\sigma^2(t)]\mu(t) - \beta\sigma^2(t) \\
   d\sigma^2/dt &=& 2D-2k\sigma^2(t) - 2\alpha\sigma^4(t) 
\end{eqnarray*}
with initial condition $\sigma^2(0)=D/k$, $\mu(0)=0$, and
$\nu(0)=0$.  We thus have
\begin{eqnarray*}
  \sigma^2(t) &=& 2D 
    \frac{(\omega+k)+(\omega-k)e^{-2\omega t}}
     {(\omega+k)^2-(\omega-k)^2e^{-2\omega t}}  \\
		\\
  \mu(t) &=& \left(\frac{2\beta D}{\omega}\right)
    \frac{-1+ e^{-\omega t}}
      {(\omega+k)-(\omega-k)e^{-\omega t}}     \\
		\\
   \nu(t) &=& \left(\frac{\beta^2D}{\omega^2}-\frac{\omega-k}{2}
         -\gamma \right)t + \frac{4\beta^2D}{\omega(\omega-k)}
	\left[\frac{1}{(\omega+k)-(\omega-k)e^{-\omega t}} 
              -\frac{1}{2k} \right]		 \\
		\\
     & & \hspace{1.6in} -\frac{1}{2}\ln\left[\frac{(\omega+k)^2
       -(\omega-k)^2e^{-2\omega t}}{4\omega k} \right]
\end{eqnarray*}
where $\omega^2=k^2+4D\alpha$.

\vskip 1.0cm \noindent
\centerline{\bf References}
\vskip 0.5cm \noindent
\def \SN{\item{}}
\small
\begin{description}

\SN$^1$ B. Sakmann and E. Neher, {\it Single-channel recording},
2nd Ed. (Plenum Press, New York, 1995)

\SN$^2$ X. S. Xie, and J. K. Trautman, Ann. Rev. Phys. Chem. {\bf 49}, 
441 (1998). 

\SN$^3$ X. S. Xie and H. P. Lu, J. Biol. Chem. {\bf 274}, 15967 (1999).

\SN$^4$ H. Qian and E. L. Elson, Biophys. J. {\bf 76}, 1598 (1999).

\SN$^5$ H. Qian, Biophys. J. {\bf 79} 137 (2000).

\SN$^6$ H. Qian, J. Math. Biol. {\bf 41}, 331 (2000).

\SN$^7$ D. A. McQuarrie, J. Appl. Prob. {\bf 4}, 413 (1967).

\SN$^8$ J. Keizer, {\it Statistical Thermodynamics of Nonequilibrium 
Processes} (Springer-Verlag, New York, 1987).

\SN$^9$ M. Doi and S. F. Edwards, {\it The Theory of Polymer
Dynamics} (Clarendon, Oxford, 1986).

\SN$^{10}$ N. Agmon, and J. J. Hopfield, J. Chem. Phys. 
{\bf 78}, 6947 (1983).

\SN$^{11}$ P. H\"{a}nggi, P. Talkner, and M. Borkovec, Rev. Mod. Phys. 
{\bf 62}, 251 (1990).

\SN$^{12}$ M. Qian, and B. Zhang, Acta Math. Appl. Sinica {\bf 1}, 
168 (1984).

\SN$^{13}$ M.-P. Qian, M. Qian, and G. L. Gong, Contemp. Math. 
{\bf 118}, 255 (1991).

\SN$^{14}$ H. Qian, Proc. R. Soc. Lond. A. {\bf }, in the press (2001). 

\SN$^{15}$ H. Qian, J. Math. Chem. {\bf 27}, 37 (2000). 

\SN$^{16}$ T. L. Hill, {\it Free Energy Transduction in Biology} 
(Academic Press, New York, 1977).

\SN$^{17}$ T. L. Hill, {\it Free Energy Transduction and Biochemical
Cycle Kinetics} (Springer-Verlag, New York, 1989).

\SN$^{18}$ C.S. Peskin, G.B. Ermentrout, and G.F. Oster, in 
{\it Cell Mechanics and Cellular Engineering}, edited by 
V.C. Mow, F., Guilak, R. Tran-Son-Tay, and R. M. Hochmuth 
(Springer-Verlag, New York, 1994), pp. 479-489. 

\SN$^{19}$ R.D. Astumian, Science {\bf 276}, 917 (1997).

\SN$^{20}$ H. Qian, Biophys. Chem. {\bf 67}, 263 (1997).

\SN$^{21}$ F. J\"{u}licher, A., Ajdari, and J. Prost, Rev. Mod. Phys. 
{\bf 69}, 1269 (1997).

\SN$^{22}$ H. Qian, Phys. Rev. Lett. {\bf 81}, 3063 (1998).

\SN$^{23}$ M. E. Fisher and A. B. Kolomeisky, Proc. Natl. Acad.
Sci. USA {\bf 96}, 6597 (1999). 

\SN$^{24}$ M. E. Fisher and A. B. Kolomeisky, Physica A {\bf 274},
241 (1999). 

\SN$^{25}$ H. Qian, Biophys. Chem. {\bf 83}, 35 (2000). 

\SN$^{26}$ D. Keller and C. Bustamante, Biophys. J. {\bf 78}, 
541 (2000).

\SN$^{27}$ J. D. Murray, {\it Mathematical Biology}, 2nd, corrected Ed.
(Springer-Verlag, New York, 1993).

\SN$^{28}$ H. Qian, and J.D. Murray, App. Math. Lett. {\bf 14},
405 (2001).

\SN$^{29}$ H. Risken, {\it The Fokker-Planck Equation: Methods of
Solution and Applications} (Springer-Verlag, New York, 1984).

\SN$^{30}$ L. Gammaitoni, P. H\"{a}nggi, P. Jung, and F. Marchesoni,
Rev. Mod. Phys. {\bf 70}, 223 (1998). 

\SN$^{31}$ H. Qian and M. Qian, Phys. Rev. Lett. {\bf 84}, 2271 (2000).

\SN$^{32}$ E. Florin, V. T. Moy, and H. E. Gaub,  Science 
{\bf 264}, 415 (1994).

\SN$^{33}$ V. T. Moy, E. Florin, and H. E. Gaub, Science 
{\bf 266}, 257 (1994).

\SN$^{34}$ A. Chilkoti, T. Boland, R. D. Ratner, and P. S. Stayton, 
Biophys. J. {\bf 69}, 2125 (1995).

\SN$^{35}$ N. G. van Kampen, {\it Stochastic Processes in 
Physics and Chemistry}, Revised and enlarged Ed.
(North-Holland, Amsterdam, 1997).

\SN$^{36}$ B. E. Shapiro and H. Qian, Biophys. Chem. {\bf 67}, 211 (1997).

\SN$^{37}$ B. E. Shapiro and H. Qian, J. Theoret. Biol.
{\bf 194}, 551 (1998).

\SN$^{38}$ H. Qian and B. E. Shapiro, Prot: Struct. Funct. Genet. 
{\bf 37}, 576-581 (1999).

\SN$^{39}$ E. Evans and K. Richie, Biophys. J. {\bf 72}, 1541 (1997).

\SN$^{40}$ E. Evans and K. Richie, Biophys. J. {\bf 76}, 2439 (1999).

\SN$^{41}$ M. S. Z. Kellermayer, S. B. Smith, H. L. Granzier, and
C. Bustamante, Science {\bf 276}, 1112 (1997).

\SN$^{42}$ M. Rief, M. Gautel, F. Oesterhelt, J. M. Fernandez, and
H. E. Gaub, Science {\bf 276}, 1109 (1997).

\SN$^{43}$ L. Tskhovrebova, J. Trinick, J. A. Sleep, and R. M.
Simmons, Nature {\bf 387}, 308 (1997).

\SN$^{44}$ M. C. Mackey, Rev. Mod. Phys. {\bf 61}, 981 (1989).

\SN$^{45}$ H. Qian, Phys. Rev. E. in the press.

\SN$^{46}$ J. Howard, Ann. Rev. Physiol. {\bf 58}, 703 (1996).

\SN$^{47}$ T. L. Hill, Progr. Biophys. Mol. Biol. {\bf 28}, 
267 (1974).

\SN$^{48}$ A. F. Huxley, Progr. Biophys. {\bf 7}, 255 (1957).

\SN$^{49}$ T. Elston, H.Y. Wang, and G. Oster, Nature 
{\bf 391}, 510 (1998)

\SN$^{50}$ H.Y. Wang, and G. Oster, Nature {\bf 396}, 279 (1998).

\SN$^{51}$ C. Blomberg, J. Mol. Liquids {\bf 42}, 1 (1989).

\SN$^{52}$ R. Zwanzig, Proc. Natl. Acad. Sci. USA, {\bf 85}, 
2029 (1988).

\SN$^{53}$ R. Zwanzig, J. Chem. Phys. {\bf 97}, 3587 (1992).

\SN$^{54}$ J. Wang and P.G. Wolynes, Chem. Phys. Lett. 
{\bf 212}, 427 (1993).

\SN$^{55}$ J. Wang and P.G. Wolynes, Chem. Phys. {\bf 180}, 141 (1994).

\SN$^{56}$ J. Wang and P.G. Wolynes, Phys. Rev. Lett. 
{\bf 74}, 4317 (1995).

\SN$^{57}$ E. Di Cera, J. Chem. Phys. {\bf 95}, 5082 (1991).

\SN$^{58}$ D. E. Koshland, Proc. Natl. Acad.  Sci. USA 
{\bf 44}, 98 (1958).

\SN$^{59}$ H. Qian and J. J. Hopfield, J. Chem. Phys. 
{\bf 105}, 9292 (1996).

\SN$^{60}$ H. Qian, J. Chem. Phys. {\bf 109}, 10015 (1998). 

\SN$^{61}$ J. Wyman, J. Mol. Biol. {\bf 11}, 631 (1965). 

\SN$^{62}$ H. M. Taylor and S. Karlin,{\it An Introduction to
Stochastic Modeling}, 3rd Ed. (Academic Press, San Diego, 1998).  

\SN$^{63}$ R. Zwanzig, Proc. Natl. Acad. Sci. U.S.A. {\bf 92}, 
9801 (1995).

\SN$^{64}$ R. Doyle, K.T. Simon, H. Qian, and D. Baker, 
Prot: Struct. Funct. Genet. {\bf 29}, 282 (1997).

\SN$^{65}$ H. Qian and S.I. Chan, J. Mol. Biol. {\bf 286}, 
607 (1999).

\end{description}

\pagebreak

\normalsize
\baselineskip = 0.29 in	

\vskip 1.0cm \noindent
\centerline{\bf Figure Captions}

\vskip 0.5cm \noindent
\underline{Figure 1.}  The force-displacement curve
for a simple Lennard-Jones bond calculated based on the
solution of stochastic dynamics Eq. (\ref{Smo}). 
The $F_{int}=1/x^7-1/x^{13}$ which equals zero when
$x=1$, and $F_{ext}(x)=k(x-vt-1)$.
The ordinate is $\langle F\rangle$ = 
$\int_0^{\infty} F_{ext}(x) P(x,t)dx$, and the
abscissa is $\langle x\rangle$.  The line labeled
LJ is the expected Lennard-Jones force.  As one can
see, with increasing temperature, decreasing retracting  
rate, and stiffer probe, the measured force-displacement
curve approaches to the LJ curve.  It is also noted
that with small $k_BT$, there is a mechanical 
``bond rupturing''$^{36,38}$, while at larger
$k_BT$, the maximal force can be significantly
smaller than that of LJ due thermally activated transtion.

\vskip 0.5cm \noindent
\underline{Figure 2.} (A) A conceptual energy landscape for a
globular protein, cooperative folding$^{63-65}$ (not to the 
appropriate scale) with a transition state at $x^{\ddag}$.  For a
single domain of titin, the reasonable $x_o = 10\AA$ and $L=30nm$.
The folded state is represented by the deep energy well and unfolded
(random coil) state is represented by a shallow well with large
entropy.  The reaction coordinate (the abscissa) is uniquely
defined by the direction of the mechanical force which pulls the
molecule.  At very small $x$, there is a closely packed core of
all the atoms in the domain.  Large $x$ asymptotically approaches
the contour length $L$ of the polypeptide chain:
\[   E(x) = \alpha\left\{\frac{V_0}{x_o}\left[
           -2\left(\frac{x_o}{x}\right)^6
           +\left(\frac{x_o}{x}\right)^{12}\right]\right\}
           - \beta x
           + \gamma \left[\frac{k_BTL}{\ell_p(1-x/L)}\right]    \]
where $x_o$ is the size of folded protein, $V_0$ is the
energy of the folded state, and $\ell_p$ is the persistence length
of the polypeptide random coil. $\alpha$, $\beta$ and $\gamma$ are
parameters characterizing folded,
unfolded, and stretched states of the molecule.  In the figure
they are chosen as $\alpha V_0/x_o = 20$, $\beta = 1$, and
$\gamma k_BTL/\ell_p = 1$, and $L = 10$.  (B) The corresponding
force as function of displacement, $F_1(x) = dE(x)/dx$.  $I$, $II$,
and $III$ are used to label the three monotonic regions of the
curve.  Note region $II$ including the transition state is mechanically
unstable.

\vskip 0.5cm \noindent
\underline{Figure 3.} The force-extension relation for a
series of three globular domains each of which is characterized
by Fig. 2B.  The integers by the curves are labels for the branches
in the curve. For three subunits in serial, the force is the same on
different subunits and the total extension is the sum of the
three individual extensions.  Each branch is a sum of three from
the $I$, $II$, and $III$ in Fig. 2B.  Therefore, there are total 27
branches for a trimer, but only 10 are distinguishable (see Table I).
Any branch involves $II$ (dotted lines) is mechanically unstable.
The stable branches are 1, 5, 8, 10.  The dashed lines with the
slope $-k$ represent the force-displacement for the elastic force
probe with stiffness $k$.$^{36,38}$  The bold sawtooth curve$^{42}$ 
is expected from a mechanical force-extension experiment.  A 
measurement using a force probe with less stiffness and slower
rate will show less of the sawtooth pattern.

\vskip 0.5cm \noindent
\underline{Figure 4.}  $x$ = $n-1$, $n$, $n+1$, ... in the figure
represent the periodic binding sites of a motor protein along 
its track.  $A$, $B$ and $C$ are the chemical states of the 
motor protein, i.e., the cyclic hydrolysis reaction can be 
written as $A\rightleftharpoons B 
\rightleftharpoons C \rightleftharpoons A$.  If the
potential energy $U_{int}(x,k)$ is such that the motor can
move along the track only simultaneously when 
$B\longrightarrow C$, and there is well-defined energy 
barriers between $x=n$ and $x=n+1$ for $A$, $B$, and $C$,  
then we have a simplified discrete model (line with bold face) 
for the stochastic kinetics of a motor protein.  One of the 
most important consequences of these assumptions is that the ATP 
hydrolysis and the motor protein stepping are tightly 
coupled.$^{25}$

\pagebreak

\centerline{Table I}
\vskip 0.3cm
\centerline{ \begin{tabular}{|l|c|c|c|c|c|c|c|c|c|c|} \hline
curve \# & 1& 2& 3& 4& 5& 6& 7& 8& 9& 10 \\ \hline
composition & I,I,I& I,I,II&  I,II,II& II,II,II& I,I,III&
I,II,III& II,II,III& I,III,III& II,III,III& III,III,III  \\ \hline
multiplicity & 1& 3& 3& 1& 3& 6& 3& 3& 3& 1 \\ \hline
\end{tabular} }

\vskip 0.75in \noindent
{\bf Table Caption}
\vskip 0.5cm \noindent
\underline{Table I} Each branch in Fig. 3 consists of a sum of
3 terms in Fig. 2B: $I$, $II$, and $III$ (Colume 2).  There are total
27 branches, but some of them overlap and Colume 3 shows the
multiplicity.

\pagebreak

\begin{figure}[h]
\[
\psfig{figure=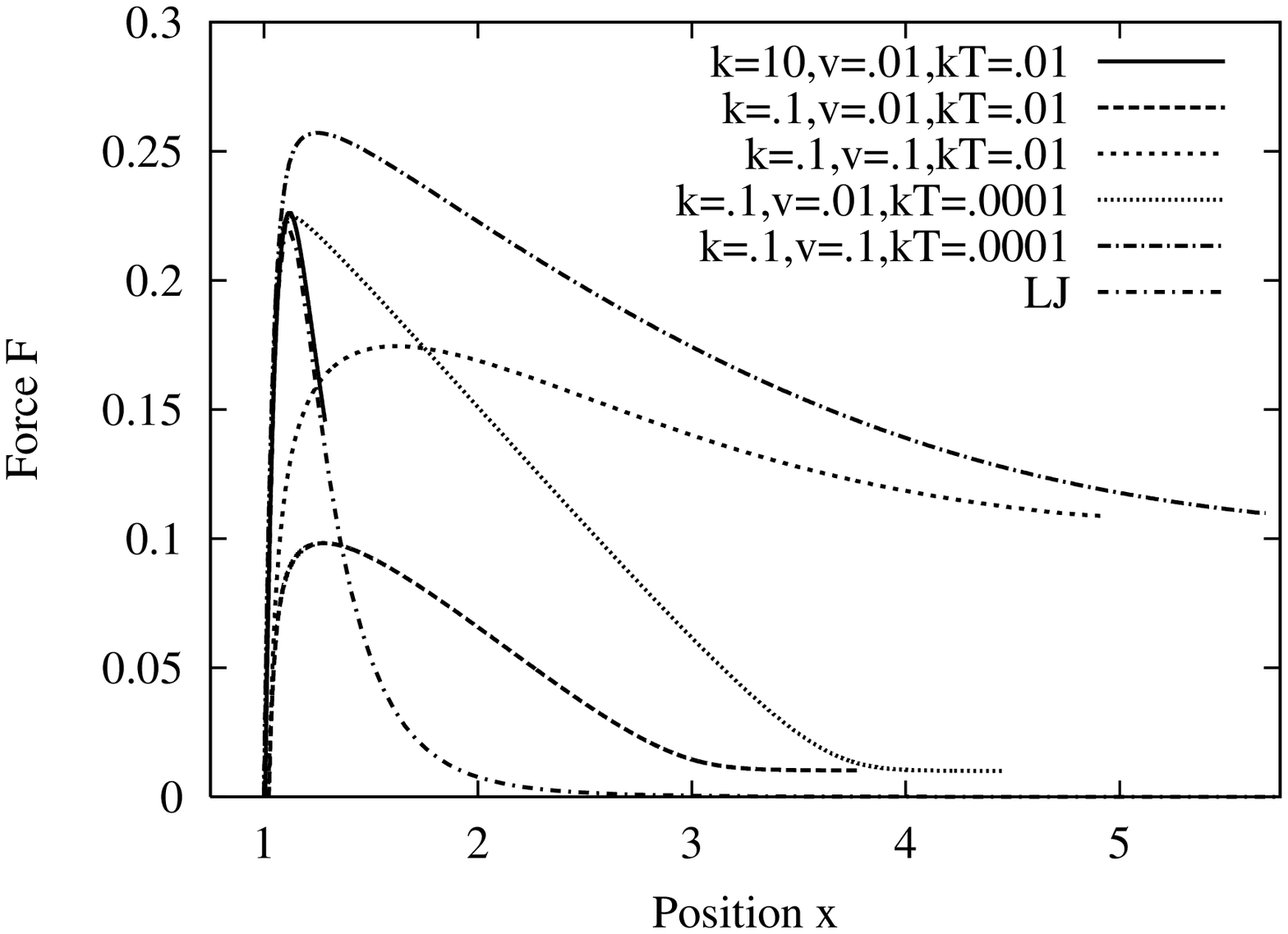,%
width=6.0in,height=3.25in,%
bbllx=2.5in,bblly=0.5in,%
bburx=7.5in,bbury=9.0in,%
angle=-90}
\]
\caption{}
\end{figure}

\pagebreak

\begin{figure}[h]
\[
\psfig{figure=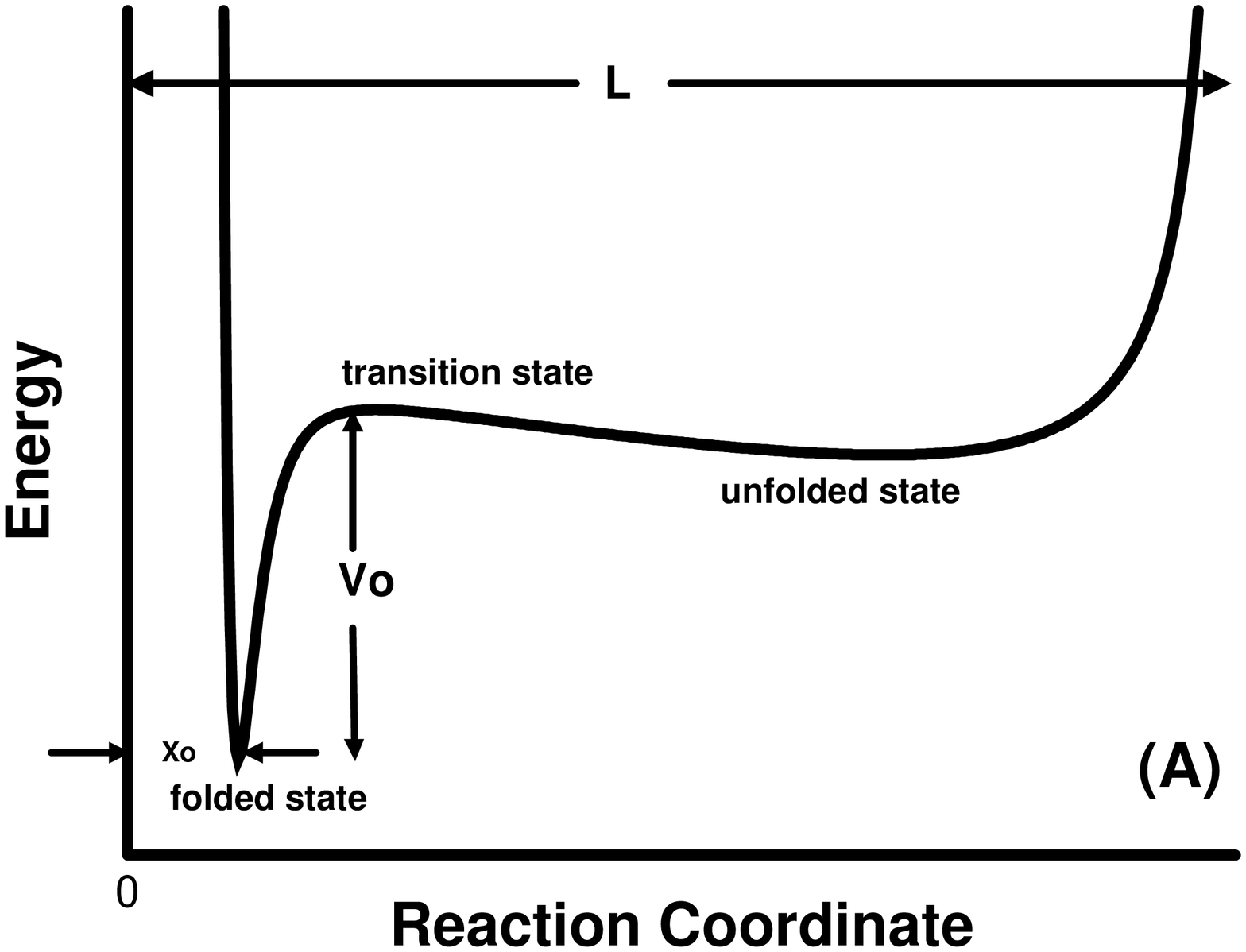,%
width=4.5in,height=3.25in,%
bbllx=1.00in,bblly=1.0in,%
bburx=7.75in,bbury=9.5in,%
angle=-90}
\]
\vskip -0.5cm
\[
\psfig{figure=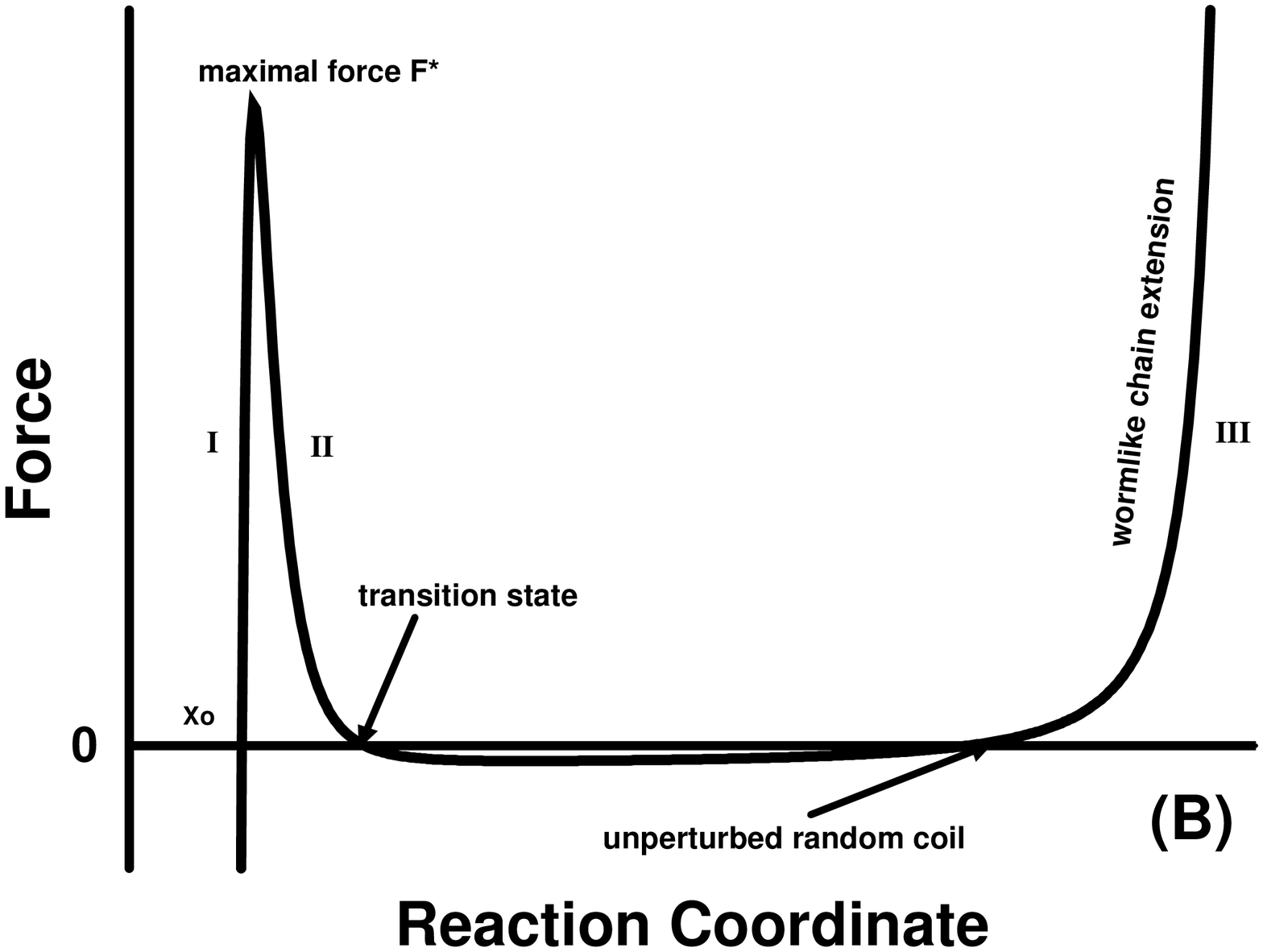,%
width=4.5in,height=3.25in,%
bbllx=1.00in,bblly=1.0in,%
bburx=7.75in,bbury=9.5in,%
angle=-90}
\]
\vskip 0.50cm
\caption{ }
\end{figure}

\pagebreak

\begin{figure}[h]
\[
\psfig{figure=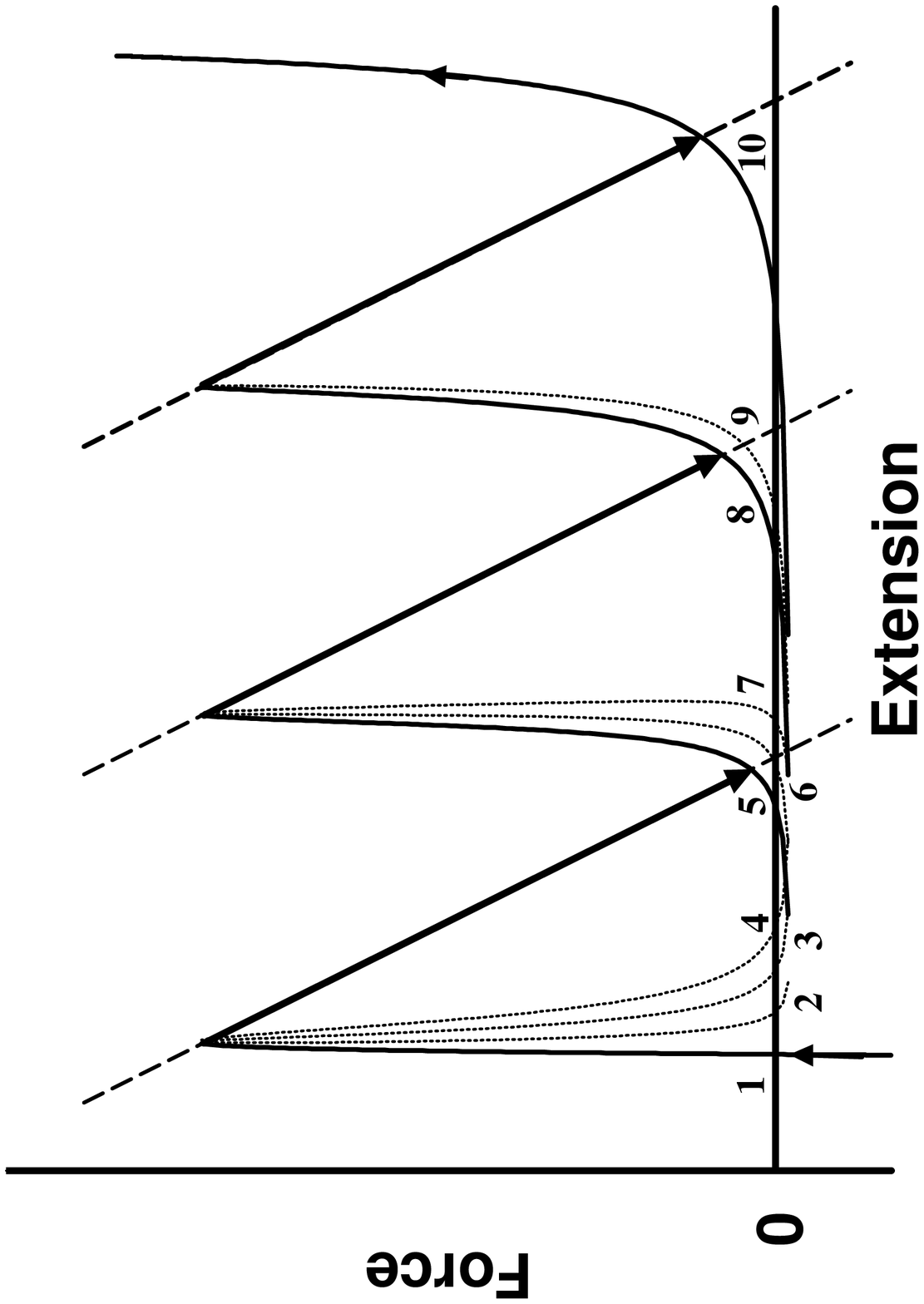,%
width=5.5in,height=6.5in,%
bbllx=1.00in,bblly=1.0in,%
bburx=7.75in,bbury=9.5in}
\]
\vskip 0.50cm
\caption{ }
\end{figure}

\pagebreak

\begin{figure}[h]
\[
\psfig{figure=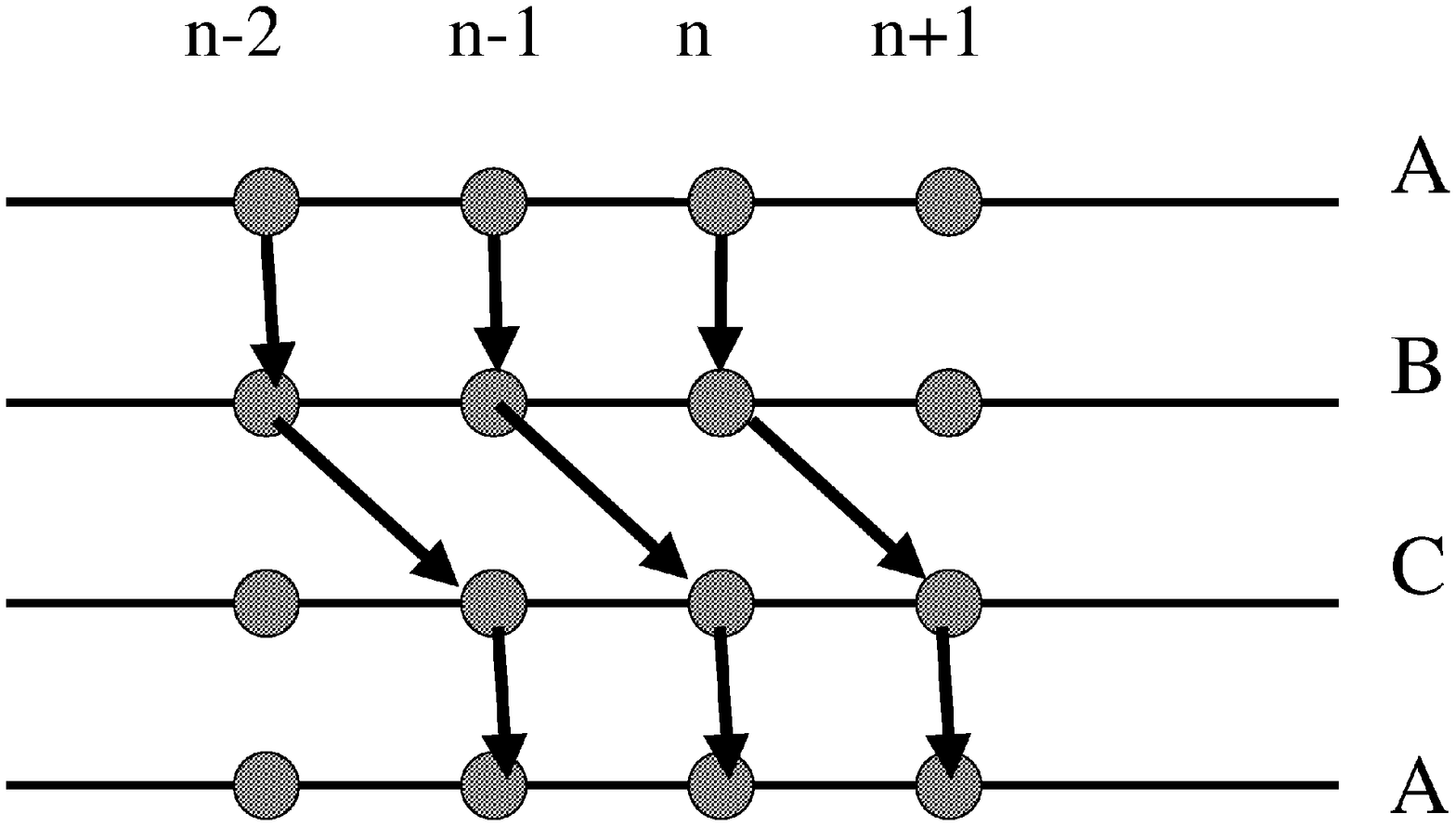,%
width=5.75in,height=4.5in,%
bbllx=1.in,bblly=1.0in,%
bburx=7.75in,bbury=10.0in,%
angle=-90}
\]
\caption{}
\end{figure}

\end{document}